\documentclass[preprint]{aastex62}
\usepackage{color}
\usepackage{multirow}
\usepackage{booktabs}
\usepackage{threeparttable}

\graphicspath{{./}{figures/}}

\shorttitle{DM of FRB host galaxy from IllustrisTNG}
\shortauthors{Zhang et al.}

\begin{document}

\title{Dispersion measures of fast radio burst host galaxies derived from IllustrisTNG simulation}

\author[0000-0001-6545-4802]{G. Q. Zhang}
\affiliation{School of Astronomy and Space Science, Nanjing University, Nanjing 210093, China}

\author[0000-0002-1840-8835]{Hai Yu}
\affiliation{Department of Astronomy, School of Physics and Astronomy, Shanghai Jiao Tong University, Shanghai, China}

\author{J. H. He}
\affiliation{School of Astronomy and Space Science, Nanjing
University, Nanjing 210093, China} \affiliation{Key Laboratory of
Modern Astronomy and Astrophysics (Nanjing University), Ministry of
Education, Nanjing 210093, China}

\author[0000-0003-4157-7714]{F. Y. Wang} \affiliation{School of Astronomy and Space Science, Nanjing
University, Nanjing 210093, China} \affiliation{Key Laboratory of
Modern Astronomy and Astrophysics (Nanjing University), Ministry of
Education, Nanjing 210093, China}

\correspondingauthor{F. Y. Wang}
\email{fayinwang@nju.edu.cn}

\begin{abstract}
We calculate the dispersion measures (DMs) contributed by host
galaxies of fast radio bursts (FRBs). Based on a few host galaxy
observations, a large sample of galaxy with similar properties to observed ones
has been selected from the IllustrisTNG simulation. They are used to compute the distributions of host galaxy
DMs for repeating and non-repeating FRBs. For repeating FRBs, we
infer the DM$ _{\mathrm{host}} $ for FRBs like FRB 121102 and FRB 180916 by assuming that the burst sites are tracing the star formation rates in host
galaxies. The median DM$_{\mathrm{host}}$ are $35 (1+z)^{1.08}$
and $96(1+z)^{0.83}$ pc cm$^{-3}$ for FRBs like FRB 121102 and FRB 180916, respectively.
In another case, the median of DM$_{\mathrm{host}}$ is about
$30 - 70$ pc cm$^{-3}$ for non-repeating FRBs in the redshift range $z=0.1-1.5$,
assuming that the burst sites are the locations of binary neutron
star mergers. In this case, the
evolution of the median DM$_{\mathrm{host}}$  can be  calculated by $33(1+z)^{0.84}$ pc cm$^{-3}$.  The distributions of DM$_{\mathrm{host}}$ of repeating and non-repeating FRBs can be well fitted with the log-normal function.
Our results can be used to infer redshifts of non-localized FRBs.
\end{abstract}


\section{Introduction}\label{sec:intro}
Fast radio bursts (FRBs) are radio transients with short duration
and high dispersion measures (DMs). After the first FRB was
discovered \citep{Lorimer2007Sci...318..777L}, a few hundred
FRBs have been observed by different telescopes to date
\citep{Thornton2013,Petroff2016PASA...33...45P,CHIME2019Natur.566..230C,Shannon2018Natur.562..386S}.
Observationally, there are two types of FRBs, repeating FRBs and
non-repeating FRBs. Till now, twenty repeating FRBs have been
published
\citep{Spitler2016Natur.531..202S,CHIME2019Natur.566..235C,2019ApJ...885L..24C,Fonseca2020arXiv200103595F}.
For comparison, most FRBs are apparently non-repeating events. The data of published FRBs can be found in the FRB catalog
\citep{Petroff2016PASA...33...45P}.

The physical origin of FRBs is still under debate \citep[for recent
reviews,
see][]{Platts2019PhR...821....1P,Cordes2019ARA&A..57..417C,Petroff2019A&ARv..27....4P}.
More recently, a radio burst from Galactic magnetar SGR 1935+2154 was discovered \citep{CHIME2020arXiv200510324T,
Bochenek2020arXiv200510828B}, which supports the conjecture that some FRBs
are produced by magnetars. The merger of compact binary is also a promising model for
non-repeating FRBs, which has been discussed by many works \citep{totani2013PASJ...65L..12T,Wang2016ApJ...822L...7W}.

The large DMs of FRBs are well beyond the contribution from the Milky
Way galaxy, which indicates an extragalactic origin. The
localization of FRB 121102 \citep{Tendulkar2017ApJ...834L...7T},
FRB 180924 \citep{Bannister2019Sci...365..565B}, FRB 190523
\citep{Ravi2019Natur.572..352R}, FRB 181112
\citep{Prochaska2019arXiv190911681P}, and FRB 180916
\citep{Marcote2020Natur.577..190M} support their cosmological
origin. Therefore, FRBs can be used as cosmological probes, such as
constraining the cosmological parameters
\citep{Gao2014ApJ...788..189G,
Zhou2014PhRvD..89j7303Z,Walters2018ApJ...856...65W,Jaroszynski2019MNRAS.484.1637J,Li2018NatCo...9.3833L},
measuring the cosmological proper distance
\citep{Yu2017A&A...606A...3Y}, constraining the baryon number
density \citep{Deng2014ApJ...783L..35D,Wei2019JCAP...09..039W,Li2019ApJ...876..146L,Macquart2020Natur.581..391M},
testing dark matter models
\citep{Munoz2016PhRvL.117i1301M,Wang2018A&A...614A..50W}, measuring Hubble parameter \citep{Wu2020ApJ...895...33W},
testing Einstein equivalent principle \citep{Wei2015PhRvL.115z1101W,Yu2018ApJ...860..173Y} and testing hydrogen
and helium reionization histories \citep{Fialkov2016JCAP...05..004F,Caleb2019MNRAS.485.2281C} .

The main difficulty in using FRBs for cosmological purpose is how
to determine their redshifts.
Theoretically, the redshift of FRB can be derived from the observed
DM, which is the integration of free electron number density along
a given line of sight. According to their origin, the observed DM can be divided into several parts:
\begin{equation}
\label{eq:DM}
\rm DM = DM_{\mathrm{MW}} + DM_{\mathrm{halo}} + DM_{\mathrm{IGM}} + \frac{DM_{\mathrm{host}} + DM_{\mathrm{source}}}{1 + z},
\end{equation}
which includes the contributions from the interstellar medium of the Milky Way
(DM$_{\rm{MW}}$), the halo of the Milky Way (DM$ _{\mathrm{halo}}
$), the intergalactic medium (IGM) (DM$_{\rm{IGM}}$), the host galaxy (DM$ _{\rm{host}}$)
and the source (DM$_{\mathrm{source}}$). DM$_{\rm{MW}}$ can
be derived from the NE2001 model
\citep{Cordes2002astro.ph..7156C} or YMW16 model
\citep{Yao2017ApJ...835...29Y}. We have a poor understanding of DM$
_{\mathrm{halo}}$. \citet{Dolag2015MNRAS.451.4277D} found the
typical value of DM$ _{\mathrm{halo}} \sim 30 $ pc cm$^{-3}$ from
numerical simulations. \citet{Prochaska2019MNRAS.485..648P}
estimated that DM$ _{\mathrm{halo}} $ is about $ 50 \sim 80 $ pc
cm$^{-3}$. The relation between DM$_{\rm{IGM}}$
and redshift $z$ has been studied by some works \citep{Ioka2003ApJ...598L..79I,Deng2014ApJ...783L..35D}, which
can be used to compute the pseudo redshifts of FRBs.
The source contribution DM$_{\mathrm{source}}$ depends on
the unclear central engine of FRBs. For example, if FRBs
are produced by the collapses of massive stars, the values of
DM$_{\mathrm{source}}$ are large \citep{Piro2016}. On the other hand,
if FRBs are produced by mergers of binary neutron stars
\citep{Wang2016ApJ...822L...7W,Zhang2020ApJ...890L..24Z}, they have
a small DM$_{\mathrm{source}}$
\citep{Margalit2019ApJ...886..110M,Wang2020}. If the central engine is known, the value of  DM$_{\mathrm{source}}$
can be derived analytically under some assumptions \citep{Piro2016,2017ApJ...847...22Y,Wang2020}. In our discussion,
this term is ignored. The value of DM$_{\mathrm{host}}$ is hard to
determine. For FRB 121102, \citet{Tendulkar2017ApJ...834L...7T}
obtained the value of host galaxy DM as $ 55 \leq
\mathrm{DM}_{\mathrm{host}} \leq 225$ pc cm$^{-3}$, which may consist of large contribution from the sources.
\citet{Marcote2020Natur.577..190M} estimated the DM$
_{\mathrm{host}} $ of FRB 180916 is less than 70 pc cm$ ^{-3} $.
For non-repeating FRBs, \citet{Bannister2019Sci...365..565B} found
DM$_{\mathrm{host}} \simeq 30-81$  pc cm$^{-3}$ for FRB 180924.

Until now, only ten FRBs have
been localized, including the Galactic FRB 200428.
Using Very Large Array (VLA),
FRB 121102 has been localized to a dwarf galaxy with low metallicity
and low star formation rate (SFR)
\citep{Chatterjee2017Natur.541...58C,Tendulkar2017ApJ...834L...7T}.
FRB 180916 is localized to a spiral galaxy with
high stellar mass and SFR \citep{Marcote2020Natur.577..190M}. For
non-repeating FRBs, \citet{Ravi2019Natur.572..352R} localized the
host galaxy of FRB 190523 through Deep Synoptic Array ten-antenna
prototype (DSA-10). Using Australian Square Kilometer Array
Pathfinder (ASKAP), the host galaxies of FRB 180924
\citep{Bannister2019Sci...365..565B} and FRB 181112
\citep{Prochaska2019arXiv190911681P} have also been determined.
Recently, four ASKAP localized FRBs have been released \citep{Macquart2020Natur.581..391M}.
The properties of these host galaxies are listed in Table \ref{tab:hostgalaxy}.

In this paper, we use the IllustrisTNG simulation to investigate the DM$ _{\mathrm{host}} $ of
FRBs. The IllustrisTNG project is a successor of the Illustris project. Using the moving
mesh code AREPO, they perform a cosmological magnetohydrodynamical simulation \citep{Pillepich2018MNRAS.473.4077P}.
Multiple physical processes that derive galaxy formation are implemented in AREPO.  Several simulations are performed
in this project. Each simulation contains a large number of simulated galaxies, and detailed information on each one of these galaxies
is provided.

This paper is organized as follows. In section
2, we introduce the IllustrisTNG simulation and give the method to
derive DM$_{\rm{host}}$. We will present the DM$_{\rm{host}}$
distributions for repeating FRBs like FRB 121102
in section 3, repeating FRBs like FRB 180916 in section 4
and non-repeating FRBs in section 5. In section 6, the effect of selection criteria
of galaxies is discussed and the redshifts of simulated FRBs are derived
based on our results. Finally, conclusions are given in section 7.

\section{Method}\label{sec:tng}
The DM in the FRB rest frame can be derived through
\begin{equation}
\label{eq:dm_int}
\mathrm{DM} = \int n_e dl,
\end{equation}
where $n_e$ is the electron number density and $dl$ is the light
path. If the electron distribution of a galaxy is known, the DM$_{\rm{host}}$ can be derived
through Equation \ref{eq:dm_int}. However, the electron
distribution of a galaxy can hardly be obtained from
observations. Therefore, the IllustrisTNG simulation is used to
derive the electron distribution.

The IllustrisTNG project is a successor of the Illustris
project. It is a large-volume, cosmological and
magnetohydrodynamical simulation. It consists of TNG50, TNG100, and
TNG300, which mean that the lengths of the simulation boxes are about
50 Mpc, 100 Mpc, and 300 Mpc, respectively. The large simulation box
can be used to study the large-scale structure of the Universe, while the small simulation
box can provide a better resolution of galaxies, which is important
when we are interested in the structure of galaxies. Therefore, the
TNG50 simulation is the best choice to derive the distribution of
DM$_{\rm{host}}$. However, as the data of TNG50 has not been
released, we choose the data of TNG100 to perform the analysis, which has been
released \citep{Nelson:2018uso,Pillepich2018MNRAS.475..648P,
Naiman2018MNRAS.477.1206N,Nelson2018MNRAS.475..624N,Marinacci2018MNRAS.480.5113M,Springel2018MNRAS.475..676S}.
The TNG100 includes three runs with different numbers of particles,
which are called TNG100-1, TNG100-2, and TNG100-3, respectively.
Among these three runs, we choose the TNG100-1, which contains the most particles.
The side length of this run is
110.7 Mpc and it contains 6,028,568,000 dark matter particles.
A good particle resolution can be achieved in this run.
The initial conditions of this run are consistent with
the Planck 2015 results \citep{Planck2016A&A...594A..13P}. The same
cosmological parameters are also used in this paper.

Employing the friends-of-friends (FoF) algorithm, the dark matter particles
in the TNG simulation are divided into different FoF halos (groups).
In each halo, the particles are separated into subhalos using the subfind
algorithm and the subhalos correspond to galaxies. For each subhalo, the
TNG project provides detailed information, such as the stellar mass,
SFR, gas metallicity, half mass radius, etc.
This information can help us to select similar galaxies to the observed FRB host galaxies.
Repeating FRBs and non-repeating FRBs may occur in different types
of galaxies. According to mass and SFR, we select galaxies that
are similar to the host galaxies of localized FRBs. The details will be shown in
the following sections. The data of each subhalo is divided into several parts,
including gas, dark matter, tracers, black holes, stars, and wind particles.
We download the gas
part of the selected subhalos. The TNG simulation uses Voronoi
tessellation to construct the geometry of finite volume. The information of
each Voronoi cell is given by the TNG simulation. According to the simulation,
the electron number density of each cell can be derived.
% For simplification, we assume that all FRBs occur in the center of Voronoi cells.

The electron number density of the non-star-forming cells can be obtained as follows.
The TNG simulation provides the information needed to calculate
the electron number density. According to the simulation, the electron number density $ n_e $ can be derived by
\begin{equation}
    \label{eq:ne}
    n_e = \eta_e X_H  \frac{\rho}{m_H}.
\end{equation}
where $\eta_e$ is the electron abundance, $ \rho $ is the mass density, $ m_H $ is the
mass of hydrogen atom, and
$ X_H $ is hydrogen abundance of each Voronoi cell. But for star-forming cells, the electron
abundance $\eta_e$ given by the TNG simulation is not reliable. In this case, we
consider the subgrid model proposed by \citet{SH2003MNRAS.339..289S}, which is also
adopted by the TNG simulation. According to this model, the multi-phase interstellar medium is
comprised of a cold cloud and hot gas. In our calculation, we assume that the cold cloud is
entirely neutral and the hot gas is totally ionized. This assumption has been used in many previous
works \citep{Pakmor2018MNRAS.481.4410P,Stevens2019MNRAS.483.5334S}. The fraction of
hot gas is calculated according to \citet{SH2003MNRAS.339..289S}.
The method has been discussed in \citet{SH2003MNRAS.339..289S} and the appendix of  \citet{Stevens2019MNRAS.483.5334S}.
Assuming that the hot gas is totally ionized, we derive the electron number
density of star-forming cells. The electron distribution in one Voronoi cell is assumed to be uniform, so each
region in one Voronoi cell shares the same electron number density. Therefore, the DM$_{\rm{host}}$ in the local frame can be written as
\begin{equation}
    \mathrm{DM}_{\rm{host}} = \int n_edl =\sum_i^N n_{e,i} \Delta l_i,
\end{equation}
where $i$ refers to the $i$th Voronoi cell, $l_i$ is the length of
the light travel in the $i$th cell, and $N$ is the total number of
the cells which the light travels through.

Because the TNG simulation does not provide a clear boundary of
subhalo, we use the following method to set the boundary of
integration. When we try to get the DM$_{\rm{host}}$
of a subhalo, all the cells in parent halo are selected to
analyze. The light of an FRB is emitted from a cell of this subhalo
and propagates in a random direction.
If the light travels into a cell which does not belong to
this subhalo, we stop the integration.

The location of FRB in the host galaxy is
important. Repeating FRBs and non-repeating FRBs may
have different locations in their host galaxies. The positions of
FRBs may strongly depend on the progenitor models. We will discuss
the positions in detail later.

The TNG simulation contains 100 snapshots at different redshifts for
each run. Among these snapshots, 20 snapshots are full and the remaining 80 are mini.
Only the full snapshots are considered in our calculation.
The current observations show that more than 90\% of FRBs have the pseudo redshifts $
z < 1.5 $. Thus, only the snapshots with $z < 1.5$ are selected to
analyze. According to these criteria, we select the snapshots with
redshifts $z = 0.1, 0.2, 0.3, 0.4, 0.5, 0.7, 1.0$ and $ 1.5 $. In
each snapshot, 1000 subhalos are selected to derive DM$
_{\mathrm{host}} $ for repeating FRBs like FRB 121102. But for non-repeating FRBs
and repeating FRBs like FRB 180916,
in order to save calculation time, 200 subhalos are selected. For
each subhalo, we simulate the positions of 500 FRBs and
each 10 random propagating directions are simulated for each FRB. The
details are discussed in the following sections. The evolution of
DM$_{\rm{host}}$ as a function of redshift is also studied.
In our calculation, we use PyTorch and GPU to accelerate the
calculation.

\section{Repeating FRBs like FRB 121102}\label{sec:repeat12}
So far, twenty repeating FRBs have been published
\citep{Spitler2014ApJ...790..101S,Spitler2016Natur.531..202S,
CHIME2019Natur.566..235C,2019ApJ...885L..24C,Fonseca2020arXiv200103595F}.
FRB 121102 is the first observed repeating FRBs. It has been
localized to a dwarf galaxy with stellar mass about $4 - 7 \times
10^{7} M_\odot$ and SFR about $0.4\, M_\odot$
yr$^{-1}$ \citep{Tendulkar2017ApJ...834L...7T}. Recently, repeating
FRB 180916 was localized to a spiral galaxy
\citep{Marcote2020Natur.577..190M}, which is dramatically different
from FRB 121102. The stellar mass of this galaxy is about $
10^{10} M_\odot$ and the SFR of the region around
FRB 180916 is greater than 0.016 $ M_\odot $ yr$^{-1}$. The
properties of these two galaxies are dramatically different,
so we will discuss them separately. In this section, we choose the host
galaxy of FRB 121102 as a typical host galaxy. The case of FRB 180916
will be discussed in the next section.
We select 1000 subhalos with the
stellar mass $1 - 50 \times 10^7 M_\odot$ and the SFR $0.1 - 0.7\,
M_\odot$ yr$^{-1}$. The total masses of the selected
galaxies are also collected, which include
the mass contributed by gas, stars, dark matter, and black holes.
Among these components, dark matter is the main contributor.
The median and the 1$ \sigma $
error of the total mass are given in Table \ref{tab:Midrepeat}.

It is difficult to determine the location of FRB in the galaxy. Some
models suggest that repeating FRBs originate from the newborn
neutron stars \citep{Lyubarsky2014MNRAS.442L...9L,Metzger2017ApJ...841...14M}.
Recently, a faint FRB 200428 has been discovered associated with a Galactic magnetar
SGR 1935+2154 \citep{CHIME2020arXiv200510324T,Bochenek2020arXiv200510828B}. It supports
the magnetar origin of FRBs. Newborn neutron stars are mainly produced by the core-collapse supernovae
and the formation rate may trace the SFR.
Therefore, we assume that the probability of FRBs occurring in a
cell is proportional to the SFR. For simplicity, we also
assume that FRBs only occur at the centers of cells. According to
this scenario, 500 positions of FRBs are simulated in each selected galaxy.
For each simulated FRB, ten propagation directions are randomly selected in
its rest frame. As a result, about 5$\times 10^6$
DM$_{\mathrm{host}}$ are derived at each redshift.
Considering the cosmological evolution, the observed
DM$_{\mathrm{host}}$ can be computed. The DM$_{\rm{host}}$ distribution at
$ z = 0.2 $ is shown in Figure \ref{fig:frb121102} as an example.
As for the DM$_{\rm{host}}$ distributions at other redshifts,
they have a similar shape but different parameters.
We only list the medians and the 1$\sigma$ errors of these distributions in Table \ref{tab:Midrepeat}.
Besides, the evolution of DM$_{\rm{host}}$ with redshifts is important.
We show the median and the 1$ \sigma $ error at each redshift in the left panel of Figure \ref{fig:median}
as the blue line.

From Figure \ref{fig:frb121102}, it is obvious that the
distribution of DM$_{\rm{host}}$ has a long tail.
Due to this reason, the median of
DM$_{\rm{host}}$ is more representative than the mean value.
\citet{Tendulkar2017ApJ...834L...7T} obtained the
DM$_{\rm{host}}$ of FRB 121102 between 55 and 225 $\mathrm{pc}\,
\mathrm{cm}^{-3}$ at redshift $z = 0.193$.  We show this DM$ _{\mathrm{host}} $
with the blue shaded region in Figure \ref{fig:frb121102}.
In our simulation, the
DM$_{\rm{host}}$ at $z = 0.2$ is
$42.81^{+50.73}_{-26.09}$ pc cm$^{-3}$, which is consistent with the observation at $1\sigma$ confidence level.
It must be noted that the value of DM$ _{\mathrm{source}} $ of FRB 121102 may be large. From observations, the rotation measure (RM) of
FRB 121102 is very large, which is about 10$ ^5 $ rad m$ ^{-2} $ \citep{Michilli2018Natur.553..182M}.  This large RM
originates from the local environment, which suggests a large DM$ _{\mathrm{source}} $. The
DM$ _{\mathrm{host}} $ given by \citet{Tendulkar2017ApJ...834L...7T} includes the contribution from the environment around the source.
Thus, the DM contributed by the host galaxy must be smaller. Meanwhile, it has been shown that
the host galaxy of FRB 121102 is atypical among nearby FRBs \citep{Li2019ApJ...884L..26L}.
The possible DM$ _{\rm{host}} $ of repeating FRB 180814 is given in their work, which
is DM$_{\rm{host}} < 80 $ pc cm$^{-3}$. This is also consistent with our results. The distribution of
DM$_{\mathrm{host}} $ has a long tail, which indicates some light
paths pass through the whole galaxy. The values of DMs along these
directions are very large. We use the log-normal function to fit the distribution of
DM$ _{\mathrm{host}} $,
\begin{equation}
	\label{eq:lognorm}
	P(x; \mu, \sigma) = \frac{1}{x\sigma \sqrt{2\pi}} \exp\left(- \frac{(\ln x - \mu)^2}{2\sigma^2}\right),
\end{equation}
where $ \mu $ and $ \sigma $ are free parameters. The mean and variance of this distribution
are $ e^\mu $ and $ e^{(2\mu+\sigma^2)}[e^{\sigma^2} - 1] $, respectively. In Figure \ref{fig:frb121102},
the red line is the best-fitting result and the fitting parameters are listed in Table \ref{tab:bestfit}.

\begin{figure}
    \centering
    \includegraphics[width=1.0\linewidth]{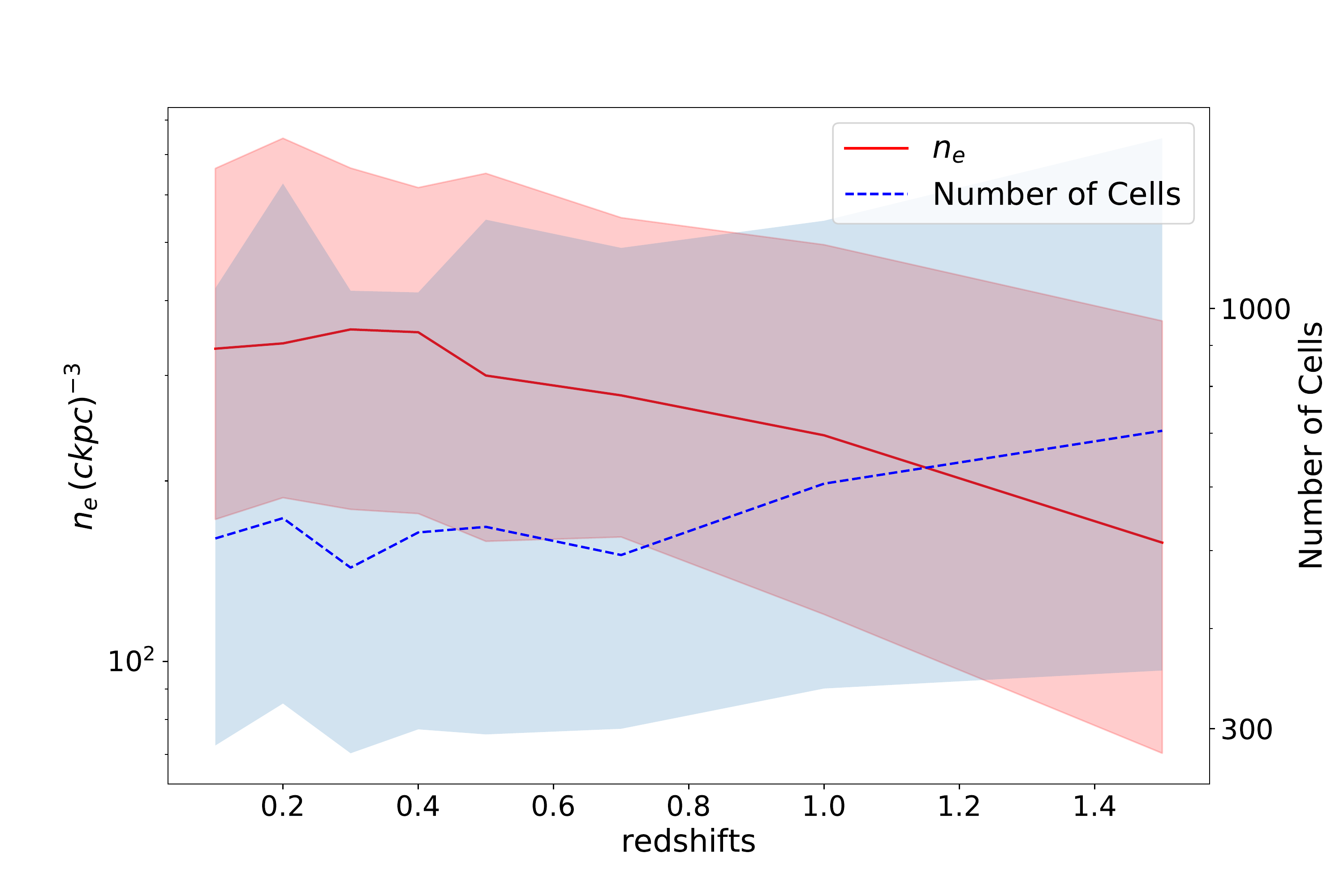}
    \caption{The comoving electron number density $ n_e $ and the number of cells in subhalos as a function of redshifts.
    These values are derived from the subhalos which are similar to the host galaxies of FRB 121102.
    The red solid line is median value of $ n_e $ with 1$ \sigma $ error (red shaded region). The blue dashed line is the
    median with 1$ \sigma $ error (blue shaded region) of the number of cells in subhalos.}
    \label{fig:repeatdecrease}
\end{figure}

We also find that DM$_{\rm{host}}$ increase as redshifts increase. The evolution
of the median of DM$_{\rm{host}}$ can be fitted by
\begin{equation}
	\label{eq:hostevolution}
	\mathrm{DM}_{\mathrm{host}}(z) = A(1+z)^\alpha.
\end{equation}
The best fitting parameters are $ A = 34.72^{+17.77}_{-14.47}$ pc cm$^{-3}$, and  $\alpha = 1.08^{+0.87}_{-0.70}$.
The fitting result is shown as the red dashed line in the left panel of Figure \ref{fig:median}.
In order to investigate this evolution, we
show the comoving electron number density at different redshifts in
Figure \ref{fig:repeatdecrease}. At each redshift, we
collect the electron number density of each cell for the selected subhalos.
The red solid line and the red shaded region are the median and 1$ \sigma $ uncertainty of the
electron number density, respectively.  The comoving electron number density decreases slowly with the increases of redshift. This
evolution is so small that we can safely neglect it. Considering the cosmological evolution,
the local electron number density at high redshift is much higher. Besides, we also
should consider the factor $(1 + z)$ when the local DM$_{\rm{host}}$ is converted to the observed
one. These effects cause the redshift evolution of DM$_{\rm{host}}$. We also count
the number of gas cells in each subhalo and show the result in
Figure \ref{fig:repeatdecrease}. The blue dashed line is the median and the blue
shaded region is the 1$\sigma $ error of the number. The number of cells remains constant for
all redshifts. Thus, it has no contribution to the DM$_{\rm{host}}$ evolution.

\begin{figure}
    \centering
    \includegraphics[width=0.8\linewidth]{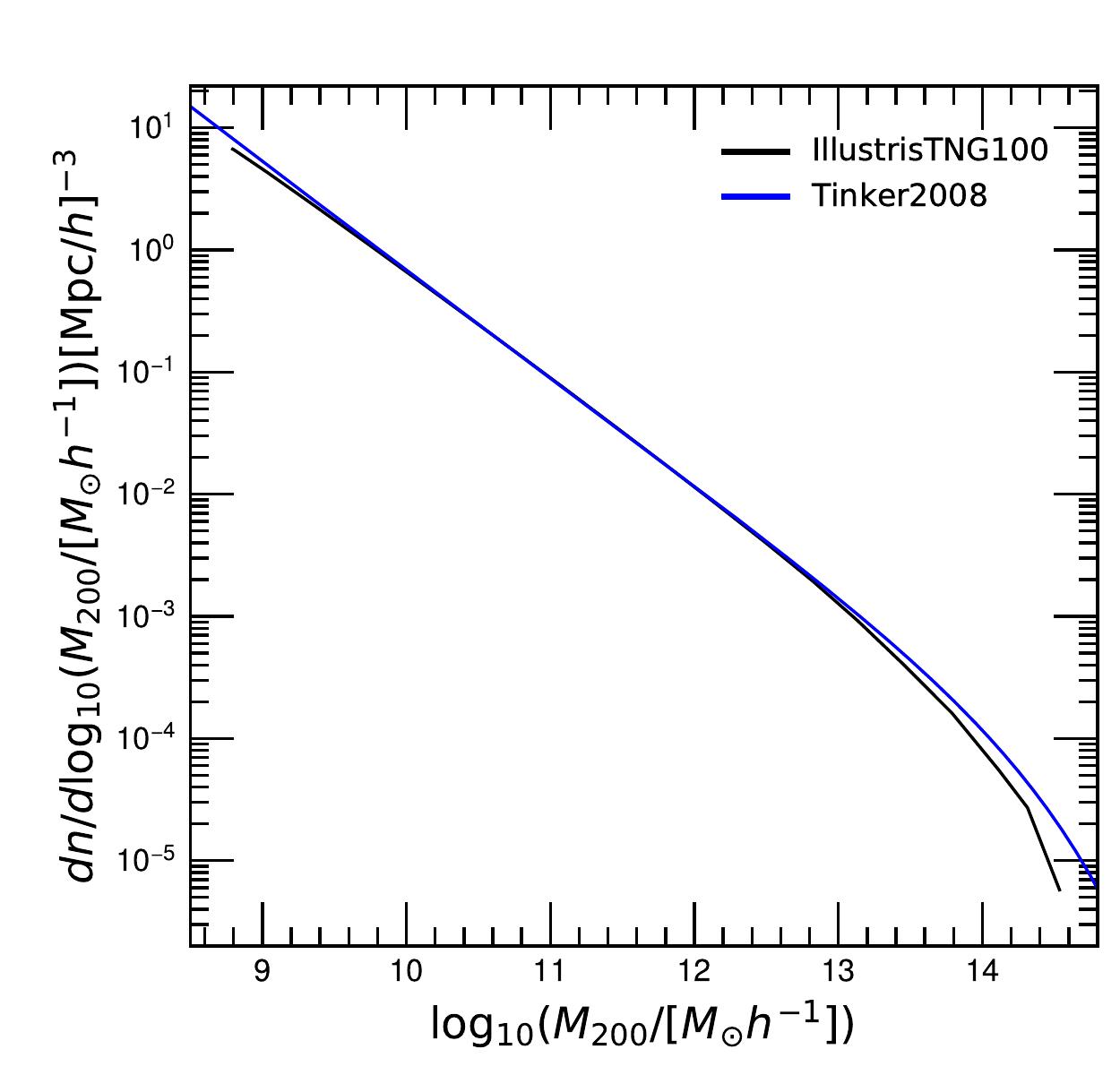}
    \caption{The halo mass function of the IllustirsTNG100 simulation versus the Tinker mass function.
    The two curves agree very well at the low-mass end, which indicates that the resolved halos in the
    IllustrisTNG100 is rather complete. The differences at the high-mass end are due to the limited box
    size of the IllustrisTNG100 simulation, which misses the most massive clusters in the universe.}
    \label{fig:ST}
\end{figure}

The mass resolution of IllustrisTNG100 is
$7.5\times10^6~M_\odot$ in dark matter particles and
$1.4\times10^6~M_\odot$ in gas particles
\footnote{https://www.tng-project.org/about/}. In our work, it is
important to test the completeness of the halo catalogue we used at
the low-mass end. To do this, we compare the measured halo mass
function from the simulation with the analytical fitting formula
proposed by Tinker \citep{Tinker:2008ff}. As the Tinker mass function
only works for a dark matter only simulation, in this test we use
the IllustrisTNG100 dark matter only run instead of the full
baryonic ones. Figure \ref{fig:ST} compares the halo mass function
of the IllustrisTNG100 simulation (black solid line) with the
Tinker's fitting formula (blue solid line). The results agree very
well at the low mass end, which indicates that the resolved halos in
the IllustrisTNG100 are complete down to $10^9M_\odot$.

\section{Repeating FRBs Like FRB 180916}\label{sec:repeat18}
The host galaxy of FRB 180916 is dramatically different
from FRB 121102. It indicates that repeating FRBs can occur in various galaxies. In this section, we investigate the distribution of
DM$ _{\mathrm{host}} $ for the host galaxies like FRB 180916.
This host galaxy is a spiral galaxy with stellar mass about $ 10^{10} M_\odot$
\citep{Marcote2020Natur.577..190M}. This repeating FRB was born in a star-forming region.
The total SFR of this galaxy is not given in their work, but they give the SFR of
a small region, which is greater than 0.016 $ M_\odot $ yr$^{-1}$.
In our analysis, the subhalos with stellar mass about
$ (0.1 - 10) \times 10^{10} M_\odot$ are selected. The total SFR
of the host galaxy is not clear, therefore, we adopt a wide range of SFR 0.01-10 $ M_\odot $ yr$^{-1}$.
In order to save the calculation time, only 200 subhalos are selected for each redshift.
Again, 500 FRB positions are simulated and 10 propagating directions are randomly selected for each FRB.
FRB 180916 is born in a star-forming region. Therefore, we assume that the probability of FRBs
borning in a cell traces the SFR of this cell. This assumption agrees with the case of repeating FRBs like FRB 121102.
The distribution of DM$ _{\mathrm{host}} $ at $ z = 0.1 $ is shown in Figure
\ref{fig:frb180916} and the medians at different redshifts are shown in the middle panel of Figure \ref{fig:median} as the blue solid line.
We also list the median and 1$ \sigma $ range at different redshifts in Table \ref{tab:Midrepeat}.
The shapes of distributions at other redshifts are similar to the distribution at $ z = 0.1 $.
The log-normal function (Equation \ref{eq:lognorm}) is used to fit the distribution of DM$ _{\mathrm{host}} $
and the best-fitting parameters are listed in Table \ref{tab:bestfit}.

In this case, the DM$ _{\mathrm{host}} $ at $ z = 0.1 $ is about 110 pc cm$^{-3}$ and
the 1$ \sigma $ region is about $ 30 - 250 $ pc cm$ ^{-3} $. \citet{Marcote2020Natur.577..190M}
estimate the DM$ _{\mathrm{host}}$ less than 70 pc cm$ ^{-3} $ for FRB 180916, which is consistent with
our results at 1$ \sigma $ level.  In Figure \ref{fig:frb180916}, the blue shaded region denotes the
DM$ _{\mathrm{host}}$ of FRB 180916.
Due to face-on view \citep{Marcote2020Natur.577..190M}, the DM$_{\rm{host}}$ of FRB 180916 is very small.
Besides, we find that the DM$ _{\mathrm{host}} $ is larger than the
results of host galaxies like FRB 121102. It is caused by the larger subhalos. The subhalos with
larger stellar mass are selected in this case. They have more cells and the light travel paths in host galaxies
are much longer. The similar evolution of DM$ _{\mathrm{host}} $ with redshifts is also found. We use Equation
(\ref{eq:hostevolution}) to fit the evolution of DM$ _{\mathrm{host}} $. The red dashed line in the middle
panel of Figure \ref{fig:median} is the best fitting with $ A = 96.22^{+50.10}_{-42.26}$ pc cm$^{-3}$, and $\alpha = 0.83^{+0.87}_{-0.58} $.
This evolution is also caused by the cosmological effect. Similar to the previous case, the galaxies
with high redshifts are much denser and the electron number densities at high redshifts are much higher.
Besides, the evolution factor $(1+z)$ needs to be considered when convert the local DM$ _{\mathrm{host}} $
to the observed one. These cosmological effects cause the evolution of DM$ _{\mathrm{host}} $.

\section{Non-Repeating FRBs}\label{sec:nonrepeat}
Recently, seven apparently non-repeating FRBs have been localized
\citep{Bannister2019Sci...365..565B,Ravi2019Natur.572..352R,Prochaska2019arXiv190911681P,Macquart2020Natur.581..391M}.
The host galaxies of these FRBs are similar. The stellar mass is in the
range $10^{9} - 10^{11} M_\odot$ and the SFR is in the range $0 - 2
M_\odot$ yr$^{-1}$. Based on these observations, we select
200 subhalos with stellar mass between $10^{9}
M_\odot$ and $2\times 10^{11} M_\odot$, and SFR in the range $0 - 2 M_\odot$
yr$^{-1}$. Similar to repeating FRBs, the locations of non-repeating
FRBs are important and difficult to determine. Their progenitor
model is different from the repeating FRBs. There are many works
suggesting that the non-repeating FRBs originate from the merger of
compact binaries
\citep[i.e.,][]{totani2013PASJ...65L..12T,Wang2016ApJ...822L...7W}.
According to these models, the positions of non-repeating FRBs
should follow the mergers of compact binaries. From observations, the positions of these FRBs are far from
the centers of host galaxies \citep{Bannister2019Sci...365..565B,Ravi2019Natur.572..352R,Prochaska2019arXiv190911681P,Macquart2020Natur.581..391M},
which also supports the conjecture that they are produced by the binary neutron star mergers. We adopt the results
of \citet{Wang2020}. Assuming that FRBs are produced by the
mergers of binary neutron stars, they calculated the locations of
FRBs in different types of galaxies through the population synthesis method \citep{Wang2020}. Because of the long
merger time of binary neutron stars and the large kick velocity, the merger locations are far from the galaxy centers.
We adopt their results for massive spiral galaxies with $10^{11} M_\odot$ case.
% alpha = 1, spiral galaxy, P 0.5, e 1
According
to their results, we simulate 500 positions of FRBs in each
galaxy. In the rest frame of each FRB, 10
random propagating directions are simulated.

For non-repeating FRBs, we show the distributions of
DM$_{\rm{host}}$ in Figure \ref{fig:nonrepeating}. The distributions at the redshifts which close to the
observed FRBs are shown. The median and 1$ \sigma $ error at all redshifts are listed in Table \ref{tab:Midrepeat}.
The long tails are shown in these distributions, which is similar
to repeating FRBs.  The log-normal function (Equation \ref{eq:lognorm})
is used to fit these distributions and the fitting results are shown as the red lines in Figure \ref{fig:nonrepeating}.
We also list the best-fitting parameters in Table \ref{tab:bestfit}.
In this case, the stellar mass
is similar to the case of repeating FRBs like FRB 180916, but the value of DM$ _{\mathrm{host}} $
is different, which is caused by the different locations. For non-repeating FRBs, we simulate
the positions according to \cite{Wang2020}. Thus, the simulated FRBs are located at the edges of the
galaxies and the DM$ _{\mathrm{host}} $ is small. Based on the observations of host galaxies, some
DM$_{\rm{host}}$ have been given in previous works \citep{Ravi2019Natur.572..352R,Bannister2019Sci...365..565B,Chittidi2020arXiv200513158C}.
We show these values in Figure \ref{fig:nonrepeating} with the blue shaded regions.
Assuming a narrow luminosity distribution,
\citet{Yang2017ApJ...839L..25Y} gave the relation between
DM$_{\mathrm{host}} $, observed flux and luminosity. They found
DM$_{\rm{host}}$ to be $267.00^{+172.53}_{-110.68}$ pc cm$^{-3}$,
which is different from our results. This may be caused by their
strong assumption on the luminosity distribution.
They assumed that the luminosity function is very narrow.
However, \citet{Luo2018} found that the luminosity function spans a
large range. This wide range of luminosity function has also been found by \cite{Agarwal2019MNRAS.490....1A}.
\citet{Li2019ApJ...884L..26L} also give similar results
for some non-repeating FRBs. In their calculation, some
possible DM$ _{\mathrm{host}} $ for the selected FRBs are given, which are
about $ 10 - 120  $ pc cm$^{-3}$.
We also show the relation between DM$_{\rm{host}}$ and
redshift in the right panel of Figure \ref{fig:median} with the blue solid line.
In this case, DM$_{\rm{host}}$ increases as redshift increases. We use equation \ref{eq:hostevolution}
to fit this evolution. The fitting result is shown in the right panel of Figure \ref{fig:median} with the red dashed line
and the best-fitting parameters are $ A = 32.97^{+23.23}_{-17.65}$ pc cm$^{-3}$ and $\alpha = 0.84^{+0.93}_{-0.60} $.
Like the previous cases, this evolution is also caused by the cosmological effects.

\section{Discussions} \label{sec:diss}

\subsection{The selection of host galaxy} \label{galaxy}
In our calculation, we select host galaxies according to
the stellar mass and the SFR. Based on the observation of FRB 121102
\citep{Tendulkar2017ApJ...834L...7T}, the galaxies with stellar mass
$1 - 50 \times 10^{7} M_\odot$ and SFR $0.1 - 0.7 M_\odot$ yr$^{-1}$ are
selected. In order to test the dependence of our results on the
properties of galaxies, we select galaxies with a narrow range of stellar mass
and SFR.

In order to compare with the observation of FRB 121102, we only select the subhalos with $z = 0.2$ to
analyze. The host galaxy of FRB 121102
is a dwarf galaxy with the stellar mass $4-7 \times 10^{7} M_\odot$ and
the SFR about $0.4 M_\odot $ yr$^{-1}$. According to this observation, we limit the stellar
mass in the range $4-7 \times 10^{7} M_\odot$ and the SFR in the range $0.1 - 0.6 M_\odot$
yr$^{-1}$. This range is narrower than we used above. In order to
save the calculation time, 200 subhalos are selected.
500 positions of FRBs are simulated and 10 propagating
directions are randomly selected. The result is shown as the blue histogram in
the top left panel of Figure \ref{fig:newRange}. For comparison, the
red histogram for the wide range case (Figure \ref{fig:frb121102}) is also shown in the top right
panel.
These two histograms are compared in the bottom panel. It's
clear that these two histograms are consistent with each other.
Therefore, if the ranges of the stellar mass and the SFR are compatible
with observations, the result is
reliable.

\subsection{Recover the Distribution of Redshifts}\label{redshift}
An important application of our results is to derive the redshifts
of non-localized FRBs. If the value of DM$_{\rm{host}}$ is known,
the redshifts of FRBs can be determined according to the relation
between DM$_{\rm{IGM}}  $ and $z$. For a large sample of FRBs without measured redshifts, the
distribution of redshifts for these FRBs can be recovered.
In this section, we use our
results to infer the pseudo redshifts of simulated FRBs and
recover the distribution of redshifts.
% Nevertheless, for repeating FRBs, the distribution
% of DM$_{\rm{host}}$ depends on redshift. It is difficult to derive
% the possible redshifts of repeating FRBs. However, for a large
% sample of repeating FRBs without measured redshifts, the
% distribution of redshifts for these FRBs can be recovered. We will
% prove it by Monte Carlo simulation.

Assuming the formation rate of FRBs traces the cosmic SFR
\citep{Madau2014ARA&A..52..415M}, we generate a mock FRB sample,
which contains 500 FRBs with redshifts. Using the relation \citep{Ioka2003ApJ...598L..79I,Deng2014ApJ...783L..35D}
\begin{equation}
\label{eq:DMIGM}
\mathrm{DM}_{\rm IGM}(z) = \frac{3cH_0\Omega_{\rm b}}{8\pi Gm_{\rm p}} f_{\rm IGM}\int_0^z \frac{H_0 f_{\rm e}(z')(1+z')}{H(z')} {\rm d} z',
\end{equation}
we can derive DM$_{\rm{IGM}}$ of these simulated FRBs. In this
equation, $f_{\rm IGM} \simeq 0.83$ is the fraction of baryon mass
in the IGM and $f_{\rm e}(z') \simeq 7/8 $ is the number ratio between the
free electrons and baryons in the IGM. We ignore
the fluctuation of DM$_{\rm{IGM}}$, which is complex and difficult
to evaluate. As for DM$_{\rm{host}}$, we use the distribution of
DM$_{\rm{host}}$ for repeating FRBs like FRB 121102 as an example.
In our results, only the distributions of DM$_{\rm{host}}$ at discrete
redshifts have been obtained. For any given redshift $z_1 < z < z_2$,
we calculate DM$_{\rm{host}}$ at $z$ using the interpolation method
\begin{equation}
\label{eq:dminter}
\mathrm{DM}_{\rm{host}}(z) = \frac{z_2 - z}{z_2 - z_1} \mathrm{DM}_{\rm{host}}(z_1) + \frac{z - z_1}{z_2 - z_1} \mathrm{DM}_{\rm{host}} (z_2),
\end{equation}
where DM$_{\rm{host}}(z_1)$ and DM$_{\rm{host}}(z_2)$ are randomly
selected from the samples which satisfies the distributions of
DM$_{\rm{host}}$ at $z_1$ and $z_2$. The Monte Carlo simulation is adopted to simulate
$\mathrm{DM}_{\rm{host}} $ of these simulated FRBs. Using this method, we simulate
the DM$_{\rm{ex}} = \mathrm{DM}_{\rm{host}} + \mathrm{DM}_{\rm{IGM}}$ of this mock sample.

The probability that an FRB with
$\mathrm{DM}_{\rm{ex}}$ located at redshift $z_1 < z < z_2$
is derived through the linear interpolation method, which is
\begin{equation}
\label{eq:redprob}
    P(\mathrm{DM}_{\rm{ex}}, z) = \frac{z_2 - z}{z_2 - z_1} P_{z_1}(\mathrm{DM}_{\rm{host}}) + \frac{z - z_1}{z_2 - z_1} P_{z_2}(\mathrm{DM}_{\rm{host}}),
\end{equation}
where $\mathrm{DM}_{\rm{host}} = \mathrm{DM}_{\rm{ex}} -
\mathrm{DM}_{\rm{IGM}}$ , $P_{z_1}$ and $P_{z_2}$ are the probability
density functions of $\mathrm{DM}_{\rm{host}}$ at redshifts $z_1$
and $z_2$, respectively. The fluctuation of DM$_{\rm{IGM}}$ is ignored, so it can be
derived from Equation \ref{eq:DMIGM}.

\begin{figure}
    \centering
    \includegraphics{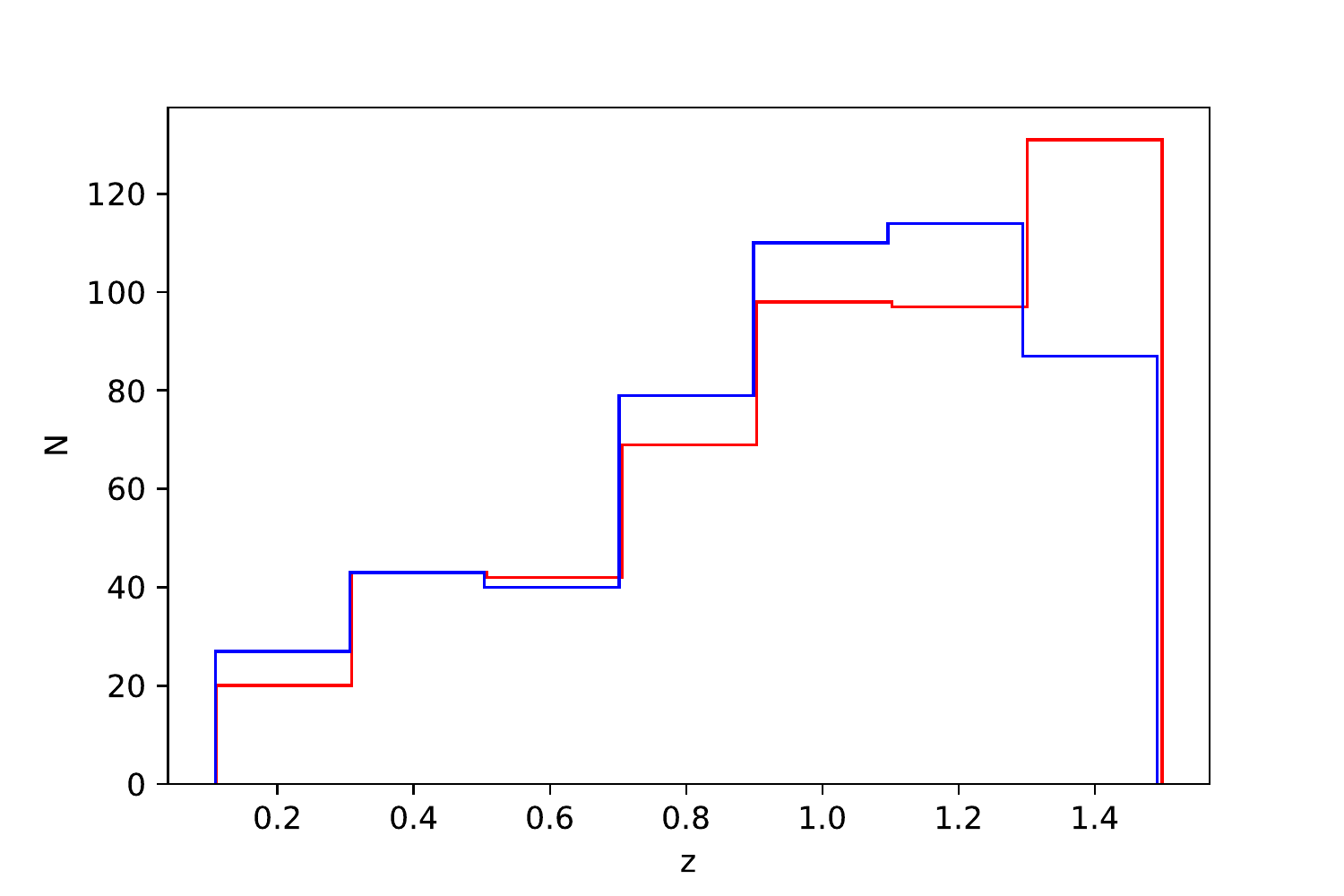}
    \caption{The histograms of simulated redshifts (blue) and derived redshifts (red) of FRBs. The two distributions are consistent with each other.}
    \label{fig:Simredshift}
\end{figure}

We use the maximum likelihood method to
derive the best-fit redshifts and show the results in Figure
\ref{fig:Simredshift}. The blue histogram is the distribution of
the simulated redshifts and the red histogram is the
derived redshifts. These two distributions are similar to each other.
Moreover, there are small
differences between the two distributions at both low redshifts and high
redshifts. The reason is that the maximum likelihood method yields
unreasonable results for those FRBs with high or extremely small
DMs. Thus, we may lose some information at low and high redshifts.

\section{Conclusions}\label{sec:conclusion}
\begin{figure}[h]
    \centering
    \includegraphics[width=0.5\linewidth]{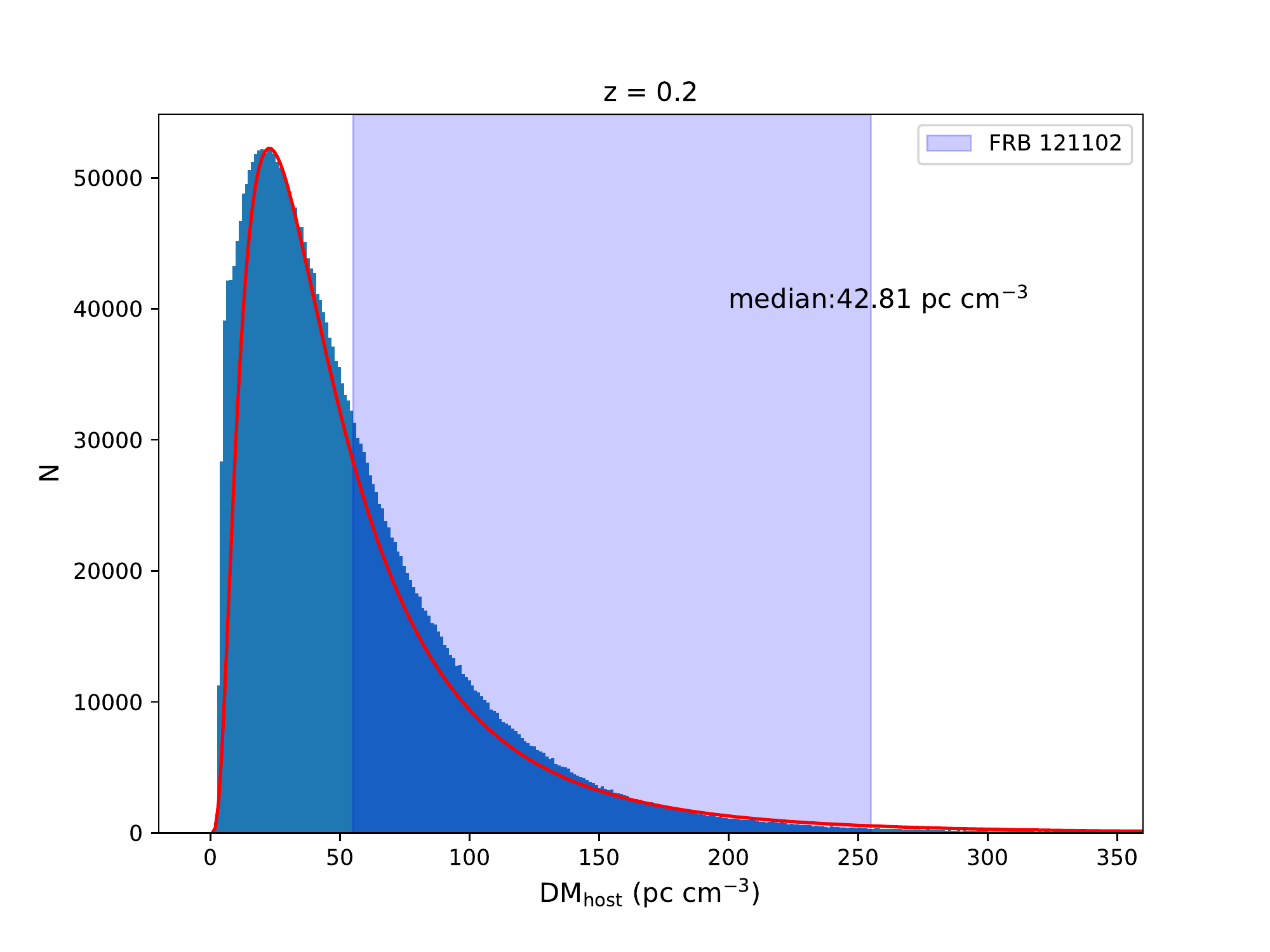}
    \caption{The distribution of $\mathrm{DM}_{\rm{host}}$ at $z = 0.2$ for repeating FRBs like FRB 121102. The red line is the best-fitting result using log-normal distribution. The best-fitting parameters are listed in
    Table \ref{tab:bestfit}. The blue shaded region is the $\mathrm{DM}_{\rm{host}}$ for FRB 121102 inferred from the observation.}
    \label{fig:frb121102}
\end{figure}

\begin{figure}[h]
	\centering
	\includegraphics[width=\linewidth]{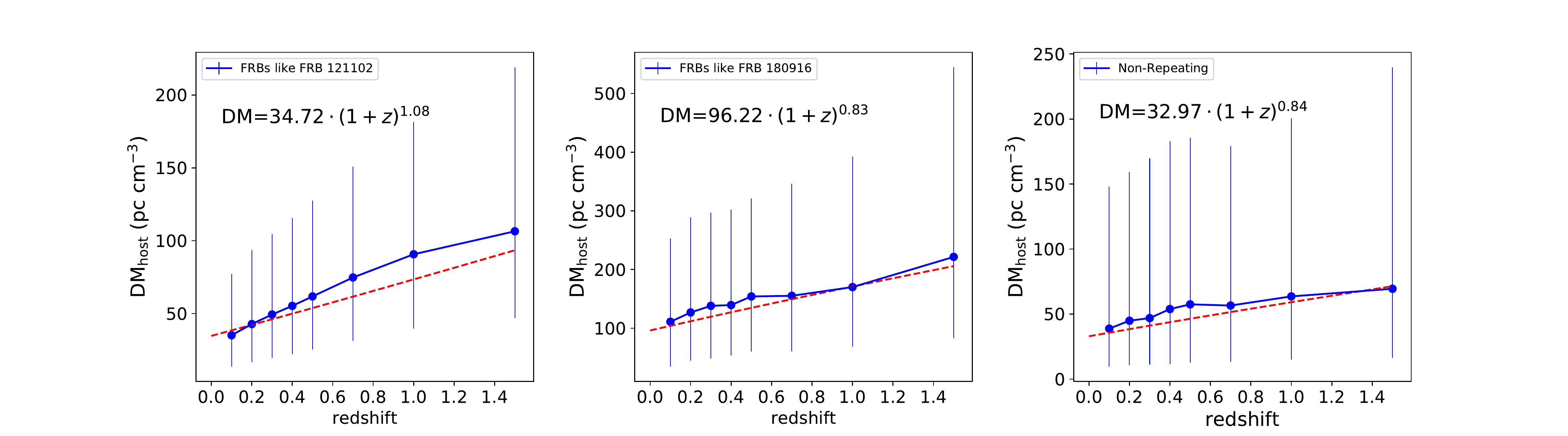}
	\caption{The median and 1$ \sigma $ error of $\mathrm{DM}_{\rm{host}}$ at different redshifts for three cases. The blue solid lines in the three panels are the evolutions of $\mathrm{DM}_{\rm{host}}$ for repeating FRBs like FRB 121102, like FRB 180916 and non-repeating FRBs, respectively. The red dashed lines are the best-fitting results using equation (\ref{eq:hostevolution}) for three cases. The fitting results are shown in the figure.}
	\label{fig:median}
\end{figure}

\begin{figure}[]
	\centering
	\includegraphics[width=0.7\linewidth]{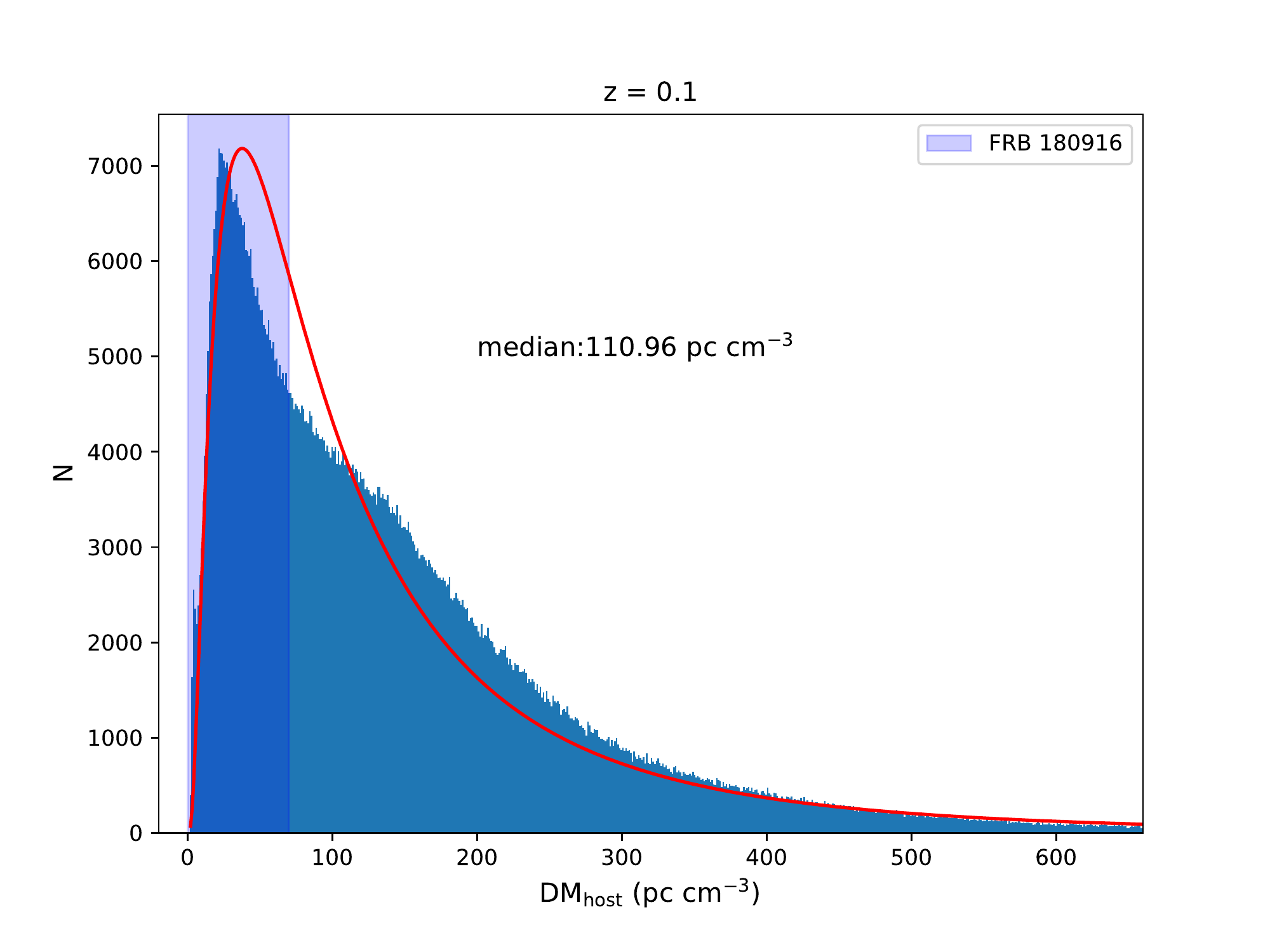}
	\caption{The distribution of $\mathrm{DM}_{\rm{host}}$ at $z = 0.1$ for repeating FRBs like FRB 180916. The red lines are the best-fitting results using log-normal distribution. The best-fitting parameters are listed in
		Table \ref{tab:bestfit}. The blue shaded region is the $\mathrm{DM}_{\rm{host}}$ for FRB 180916 inferred from the observation.}
	\label{fig:frb180916}
\end{figure}

\begin{figure}
    \centering
    \includegraphics[width=1.0\linewidth]{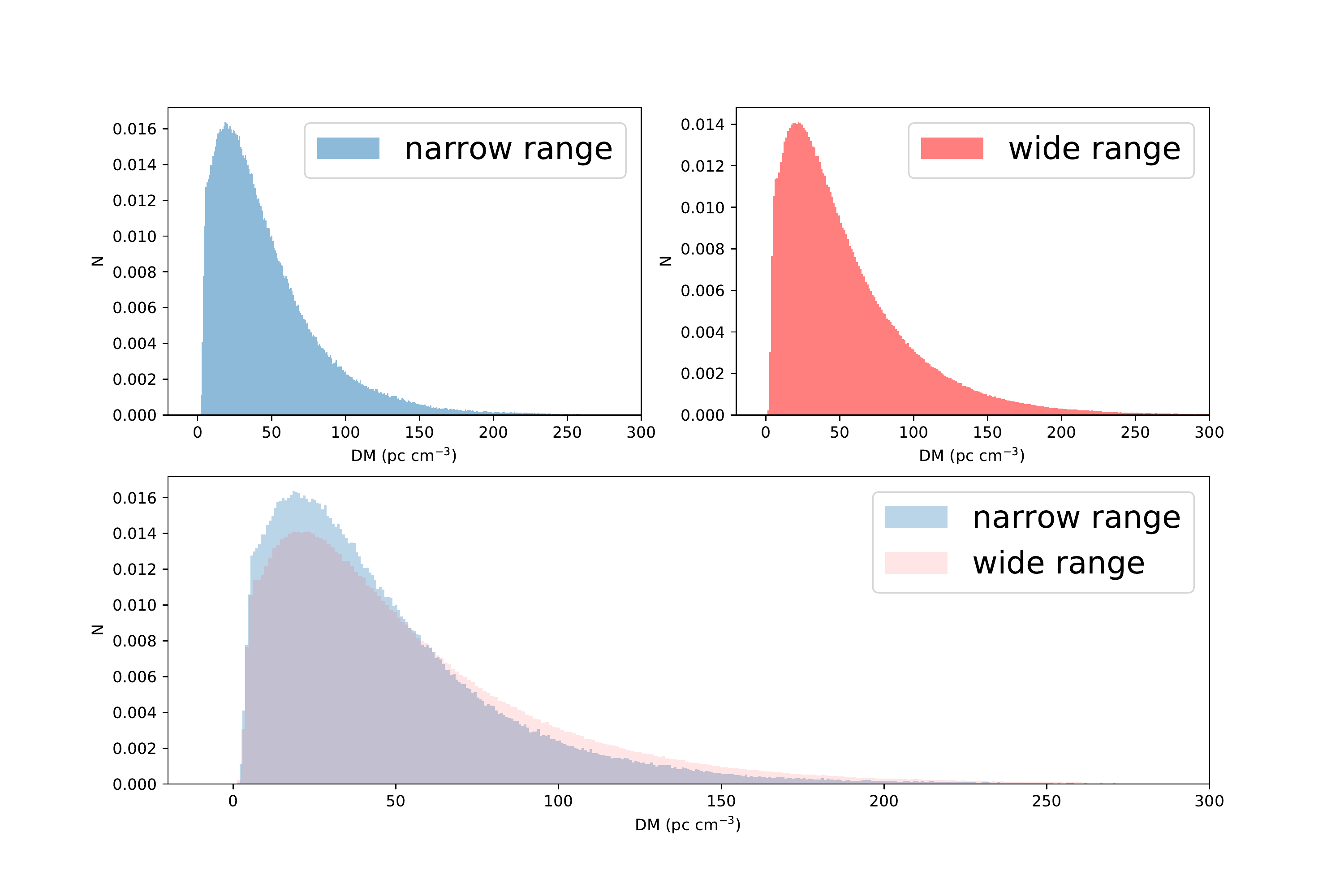}
    \caption{The distribution of DM$_{\mathrm{host}}$ for repeating FRBs at $z = 0.2$ for different stellar masses and SFRs. The blue
    histogram in the top-left panel is derived from 200 galaxies with the stellar mass $4 - 7 \times 10^7
    M_\odot$ and SFR $0.1 - 0.6 M_\odot $ yr$^{-1}$. The red histogram in the top-right panel is the previous results shown in Figure \ref{fig:frb121102}.
    We also plot these two histograms
    in the bottom panel. These two histograms are consistent
    with each other. }
    \label{fig:newRange}
\end{figure}

In this paper, we use the IllustrisTNG simulation to derive the
distributions of DMs contributed by the host galaxies of FRBs.
Based on the observations of localized FRBs, we select galaxies
to calculate the host galaxy DMs of FRBs. For
repeating FRBs like FRB 121102, we select the galaxies with stellar mass in the
range $1 - 50 \times 10^7 M_\odot$ and SFR in the range $0.1 - 0.7\,
M_\odot$ yr$^{-1}$. Assuming the formation rate of repeating FRBs in
a given cell is proportional to the SFR of this cell, 500 positions of
FRBs are simulated in each galaxy. For each FRB, ten random propagating
directions are selected. We derive the distributions of
$\mathrm{DM}_{\rm{host}}$ at different redshifts and show the result at
z = 0.2 in Figure \ref{fig:frb121102}. The observed $\mathrm{DM}_{\rm{host}}$ of FRB 121102
is also shown in Figure \ref{fig:frb121102} with the blue shaded region. In addition,
the evolution of the median with redshifts is shown in Figure \ref{fig:median}.
We find that the value of $\mathrm{DM}_{\rm{host}}$ increases as
redshift increases. We also select subhalos according to the host galaxy of repeating FRB 180916.
The host galaxy of this FRB is dramatically different from FRB 121102. Therefore,
we analyze it separately. The result is shown in Figure \ref{fig:frb180916} and
we compare our results with the observation. In this case, the $\mathrm{DM}_{\rm{host}}$ also increases as
redshift increases, which is caused by the cosmological evolution.

\begin{figure}[]
	\centering
	\includegraphics[width=1.0\linewidth]{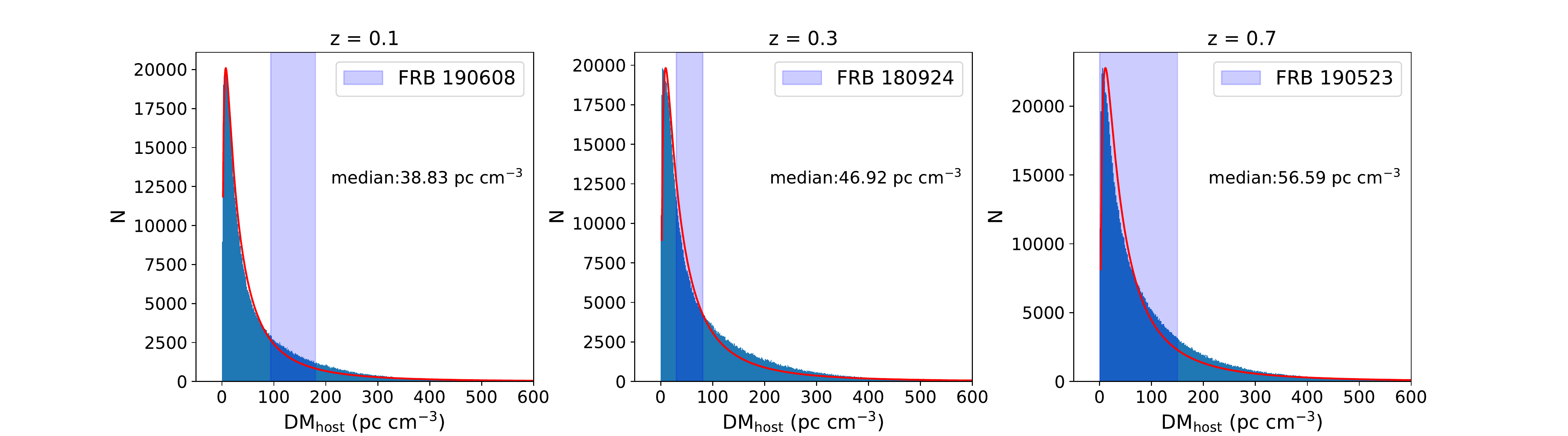}
	\caption{The distributions of $\mathrm{DM}_{\rm{host}}$ for non-repeating FRBs at $z=0.1, 0.3$, and $0.7$. The red lines are the best-fitting results using log-normal distribution and the best-fitting parameters are listed in table \ref{tab:bestfit}. The blue shaded regions are the $\mathrm{DM}_{\rm{host}}$  inferred from observations.}
	\label{fig:nonrepeating}
\end{figure}

For non-repeating FRBs, we select the galaxies with masses in the
range $10^{9} - 2 \times 10^{11} M_\odot$ and SFRs in the range $0 - 2
M_\odot$ yr$^{-1}$, which are similar to the observed non-repeating FRB host
galaxies. Assuming that the burst sites are the locations of binary
neutron star mergers, we generate 500 FRBs positions in each galaxy.
The distributions of $\mathrm{DM}_{\rm{host}}$ are shown in Figure \ref{fig:nonrepeating} and listed
in Table \ref{tab:Midrepeat}. The value of $\mathrm{DM}_{\rm{host}}$ increases as redshift increases,
 which is caused by the cosmological evolution.

The effect of the selection criteria for galaxies is also discussed. We use a narrow range of the stellar mass
and the SFR to investigate the distribution of DM$ _{\mathrm{host}} $.
The distribution in this case is consistent with the results of the
wide-parameters case. Thus, if the ranges of the stellar mass and the SFR
are compatible with the observation, the width of parameter range would
have little effect on the results. We also
use our results to infer the pseudo redshifts of simulated FRBs and show
that we can recover the distribution of simulated redshifts.
% Although the
% $\mathrm{DM}_{\rm{host}}$ of repeating FRBs depends on redshift, we
% show that the distribution of redshifts can be recovered using our
% results. Interestingly, the evolution of $\mathrm{DM}_{\rm{host}}$
% at low redshifts is unimportant. Thus, we can ignore this evolution
% for FRBs with low DMs.

% \subsection{Using FRBs as Cosmological Probes}\label{subsec:cosmo}
% Due to the high energy and cosmological origin, FRBs can be used as cosmological probes.
% But the scatter of $DM_{\rm{host}}$ and $DM_{\rm{IGM}}$ may have strong influence on the
% determination of redshifts. In the below, we will try to use FRBs to constrain the
% cosmological parameters and talk about the effects of the $DM_{\rm{host}}$.

% We will use non-repeating FRBs to analysis. According to our results, the distribution
% of $DM_{\rm{host}}$ for non-repeating FRBs
\acknowledgements We thank the anonymous referee for constructive
and helpful comments. This work is supported by the National Natural
Science Foundation of China (grant U1831207). Hai Yu is supported
by Initiative Postdocs Supporting Program (No. BX20190206).

\bibliographystyle{aasjournal}
\bibliography{ms}

\clearpage

% \usepackage{multirow}
\begin{table}[]
	\centering
	\caption{The properties of the known host galaxies of FRBs.}
	\begin{tabular}{|l|l|l|l|l|l|}
		\hline
		& FRB name   & redshift & stellar mass ($M_\odot$)       & SFR  ($M_\odot$ yr$^{-1}$)          & DM$_{\rm host}$ (pc cm$^{-3}$ )         \\ \hline
		\multirow{2}{*}{Repeating}     & FRB 121102 & 0.19      & $4-7 \times 10^7$  & 0.4            & 55-225         \\ \cline{2-6}
		& FRB 180916 & 0.03      & $\sim 10^{10}$     & 0.016\tablenotemark{*}          & \textless{}70  \\ \hline
		\multirow{7}{*}{Non-Repeating} & FRB 190523 & 0.66      & $10^{11.07}$       & \textless{}1.3 & \textless{}150 \\ \cline{2-6}
		& FRB 180924 & 0.32      & $2.2\times10^{10}$ & \textless{}2   & 30-81          \\ \cline{2-6}
		& FRB 181112 & 0.48      & $2.6\times10^{9}$  & 0.6            & -              \\ \cline{2-6}
		& FRB 190102 & 0.29      & $10^{9.5}$         & 1.5            & -              \\ \cline{2-6}
		& FRB 190608 & 0.12      & $10^{10.4}$        & 1.2            & $137\pm43$     \\ \cline{2-6}
		& FRB 190611 & 0.38      & -                  & -              &  -             \\ \cline{2-6}
		& FRB 190711 & 0.52      & -                  & -              &  -             \\ \hline
	\end{tabular}
	\tablenotetext{*}{This is the SFR of the region where FRB 180916 was born. It corresponds to a region of 1.5 kpc$ ^2 $.}
	\label{tab:hostgalaxy}
\end{table}

\begin{table}[]
    \centering
    \begin{tabular}{|c|c|c|c|c|}
        \hline
        Redshifts &  $\mathrm{DM}_{\rm{host}}$ (pc cm$^{-3}$)  & total mass  							 & $\mathrm{DM}_{\rm{host}}$ (pc cm$^{-3}$ ) &$\mathrm{DM}_{\rm{host}}$ (pc cm$^{-3}$)  \\
                  &  (like FRB 121102)    		 &   $\times 10^{10} M_\odot$, repeating & (like FRB 180916)            &(non-repeating) \\
        \hline
0.1 & $35.14^{+42.19}_{-21.46}$  & $0.09^{+0.11}_{-0.03}$ & $110.96^{+141.89}_{-76.27}$ & $38.83^{+109.34}_{-29.16}$   \\
0.2 & $42.81^{+50.73}_{-26.09}$  & $0.10^{+1.89}_{-0.04}$ & $126.84^{+162.08}_{-82.01}$ & $44.92^{+114.43}_{-34.16}$   \\
0.3 & $49.35^{+55.15}_{-29.62}$  & $0.08^{+0.15}_{-0.03}$ & $138.13^{+158.63}_{-89.57}$ & $46.92^{+122.82}_{-35.55}$   \\
0.4 & $55.29^{+60.37}_{-32.94}$  & $0.09^{+0.61}_{-0.04}$ & $139.44^{+162.88}_{-85.78}$ & $53.87^{+129.07}_{-42.34}$   \\
0.5 & $61.79^{+65.64}_{-36.46}$  & $0.10^{+0.53}_{-0.05}$ & $154.09^{+167.26}_{-93.85}$ & $57.49^{+128.12}_{-44.87}$   \\
0.7 & $74.77^{+76.07}_{-43.35}$  & $0.09^{+1.82}_{-0.03}$ & $155.24^{+190.64}_{-94.75}$ & $56.59^{+122.63}_{-43.12}$   \\
1   & $90.70^{+90.79}_{-50.87}$  & $0.12^{+3.49}_{-0.06}$ & $170.07^{+222.19}_{-101.80}$& $63.60^{+137.01}_{-48.59}$   \\
1.5 & $106.48^{+112.50}_{-59.62}$& $0.33^{+3.20}_{-0.26}$ & $221.51^{+323.08}_{-138.54}$& $69.47^{+170.33}_{-53.15}$   \\
        \hline
    \end{tabular}
    \caption{The median $\mathrm{DM}_{\rm{host}}$ with 1$\sigma$ uncertainty at different redshifts for repeating and non-repeating FRBs. We
        also give the total masses (median and 1$\sigma$ uncertainty) of the selected galaxies for repeating FRBs like FRB 121102.}
    \label{tab:Midrepeat}
\end{table}

% \usepackage{multirow}
\begin{table}[]
	\centering
	\begin{tabular}{|l|l|l|l|l|l|l|}
		\hline
		\multirow{2}{*}{redshifts} & \multicolumn{2}{l|}{\begin{tabular}{l}Repeating\\ Like FRB 121102\end{tabular}} & \multicolumn{2}{l|}{\begin{tabular}[c]{@{}l@{}}Repeating\\ Like FRB 180916\end{tabular}} & \multicolumn{2}{l|}{Non-Repeating} \\ \cline{2-7}
		& $e^\mu$ (pc cm$^{-3}$ )                                    & $\sigma$                                    & $e^\mu$ (pc cm$^{-3}$ )                                   & $\sigma$                                    & $e^\mu$  (pc cm$^{-3}$ )       & $\sigma$         \\ \hline
		0.1                        & 35.33                                      & 0.82                                        & 96.37                                      & 0.97                                        & 36.55           & 1.27             \\ \hline
		0.2                        & 43.07                                      & 0.80                                        & 121.17                                     & 0.89                                        & 40.10           & 1.25             \\ \hline
		0.3                        & 49.72                                      & 0.76                                        & 127.16                                     & 0.86                                        & 42.30           & 1.23             \\ \hline
		0.4                        & 55.55                                      & 0.76                                        & 140.58                                     & 0.80                                        & 43.60           & 1.29             \\ \hline
		0.5                        & 62.18                                      & 0.75                                        & 155.57                                     & 0.82                                        & 47.47           & 1.29             \\ \hline
		0.7                        & 75.17                                      & 0.73                                        & 157.68                                     & 0.84                                        & 48.44           & 1.21             \\ \hline
		1.0                        & 90.77                                      & 0.73                                        & 159.56                                     & 0.84                                        & 53.87           & 1.20             \\ \hline
		1.5                        & 105.74                                     & 0.75                                        & 218.74                                     & 0.90                                        & 60.98           & 1.24             \\ \hline
	\end{tabular}
	\caption{The best-fitting parameters with log-normal function of DM$ _{\mathrm{host}} $ for different cases.}
	\label{tab:bestfit}
\end{table}

% TODO: \usepackage{graphicx} required

% \begin{figure}
%     \centering
%     \includegraphics[width = 0.8\textwidth]{repeating.pdf}
%     \caption{The distributions of $\mathrm{DM}_{\rm{host}}$ at different redshifts for repeating FRBs like FRB 121102. The red lines are the best-fitting results
%     	using log-normal distribution and the best-fitting parameters are listed in Table \ref{tab:bestfit}. The blue shaded region in z = 0.2 panel is the DM$ _{\mathrm{host}} $ for FRB 121102. The distribution has a long tail,
%     which indicates some light paths pass through the whole galaxy. }
%     \label{fig:Disrepeat}
% \end{figure}
%
% \begin{figure}
%     \centering
%     \includegraphics{repeatingMedian.pdf}
%     \caption{The median $\mathrm{DM}_{\rm{host}}$ with 1$\sigma$ error at different redshifts for repeating FRBs like FRB 121102. We use $ \mathrm{DM}(z) = A(1 + z)^\alpha $ to fit the evolution of the median value. The red line is the best-fitting result with $ A =  34.72^{+17.77}_{-14.47}$ pc cm$^{-3}$ and $ \alpha = 1.08^{+0.87}_{-0.70}$.}
%     \label{fig:Midrepeat}
% \end{figure}

\end{document}